# A case study in adaptable and reusable infrastructure at the Keck Observatory Archive: VO interfaces, moving targets, and more.

G. Bruce Berriman*[a], Richard W. Cohen[b], Andrew Colson[b], Christopher R. Gelino[a], John C. Good[a], Mihseh Kong[a], Anastasia C. Laity[a], Jeffrey A. Mader[b], Melanie A. Swain[a], Hien D. Tran[b], Shin-Ywan Wang[a]

[a] NASA Exoplanet Science Institute, Infrared Processing and Analysis Center, California Institute of Technology, Pasadena, CA 91125; [b] W. M. Keck Observatory, 65-1120 Mamalahoa Hwy, Kamuela, HI 96743.

**ABSTRACT**

The Keck Observatory Archive (KOA) (https://koa.ipac.caltech.edu) curates all observations acquired at the W. M. Keck Observatory (WMKO) since it began operations in 1994, including data from eight active instruments and two decommissioned instruments. The archive is a collaboration between WMKO and the NASA Exoplanet Science Institute (NExScI). Since its inception in 2004, the science information system used at KOA has adopted an architectural approach that emphasizes software re-use and adaptability. This paper describes how KOA is currently leveraging and extending open source software components to develop new services and to support delivery of a complete set of instrument metadata, which will enable more sophisticated and extensive queries than currently possible.

In August 2015, KOA deployed a program interface to discover public data from all instruments equipped with an imaging mode. The interface complies with version 2 of the Simple Imaging Access Protocol (SIAP), under development by the International Virtual Observatory Alliance (IVOA), which defines a standard mechanism for discovering images through spatial queries. The heart of the KOA service is an R-tree-based, database-indexing mechanism prototyped by the Virtual Astronomical Observatory (VAO) and further developed by the Montage Image Mosaic project, designed to provide fast access to large imaging data sets as a first step in creating wide-area image mosaics (such as mosaics of subsets of the 4.7 million images of the SDSS DR9 release). The KOA service uses the results of the spatial R-tree search to create an SQLite data database for further relational filtering. The service uses a JSON configuration file to describe the association between instrument parameters and the service query parameters, and to make it applicable beyond the Keck instruments.

The images generated at the Keck telescope usually do not encode the image footprints as WCS fields in the FITS file headers. Because SIAP searches are spatial, much of the effort in developing the program interface involved processing the instrument and telescope parameters to understand how accurately we can derive the WCS information for each instrument. This knowledge is now being fed back into the KOA databases as part of a program to include complete metadata information for all imaging observations.

The R-tree program was itself extended to support temporal (in addition to spatial) indexing, in response to requests from the planetary science community for a search engine to discover observations of Solar System objects. With this 3D-indexing scheme, the service performs very fast time and spatial matches between the target ephemerides, obtained from the JPL SPICE service. Our experiments indicate these matches can be more than 100 times faster than when separating temporal and spatial searches. Images of the tracks of the moving targets, overlaid with the image footprints, are computed with a new command-line visualization tool, mViewer, released with the Montage distribution. The service is currently in test and will be released in late summer 2016.

**Keywords:** Archives, ground-based telescopes, software development, data management, metrics, software architecture.

---

* gbb@ipac.caltech.edu; phone 1-626-395-1817; https://koa.ipac.caltech.edu

# 1. INTRODUCTION TO THE KECK OBSERVATORY ARCHIVE

The Keck Observatory Archive (KOA)[†], a collaboration between the W. M. Keck Observatory (WMKO) and the NASA Exoplanet Science Institute (NExScI), has been in operation since 2004[1]. KOA serves the observations acquired by all 10 instruments that have been in use since the Observatory began operating in 1994, including eight active instruments and two decommissioned instruments. Newly acquired data are prepared for the archive at WMKO, and then sent electronically for ingestion into the archive at NExScI, usually by the afternoon following the observations. Principal Investigators (PIs) are guaranteed exclusive access to their data for at least 18 months following the date of observations.

Access to data is through web-based search query forms. Depending on the request, these query forms return a tabulation of all science data meeting the input criteria, with links to the FITS header, to raw (level 0) data, and associated calibration data; and, for the OSIRIS, HIRES, NIRC2 and LWS instruments, calibrated quick-look (level 1) data. Interactive exploration of these calibrated data is also offered. Data may be downloaded one-by-one or packaged in a "tar file."

As of May 1, 2016, the community has downloaded approximately 30 TB of data from the archive, and there have been 134 peer-reviewed papers published that cite KOA. Figure 1 shows the yearly growth in citation rates.

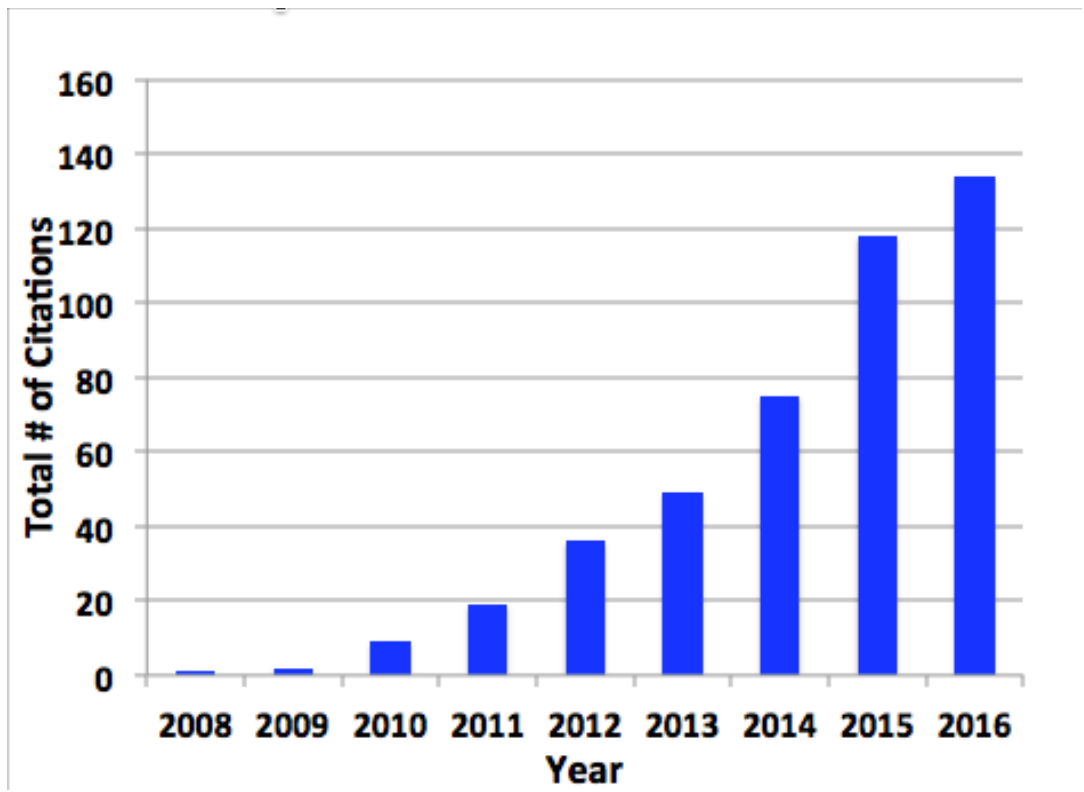

Figure 1: Cumulative number of peer-reviewed papers that cite the Keck Observatory Archive from 2008 to 2016 May 1.

The 43 papers that cited KOA in 2015 represent 12% of the scientific output of the Observatory. Despite this take-up, KOA has lacked a Virtual Observatory (VO)-compliant, program-based data discovery and access interface and the capability to discover Solar System objects, and at the recommendation of the user community, we are rectifying this state of affairs.

Our architectural approach to new services is to re-use and extend existing software as part of a component-based approach that KOA has emphasized since the archive began operations. In particular, we exploit components of the Montage Image Mosaic Engine[‡2], an open-source toolkit for building image mosaics. In particular, we are taking advantage of components that accept input from the command-line to discover the spatial-coverage of images at scale and create annotated images with overlays. We will describe the program interface first, and then the Solar System service. Throughout, we emphasize the challenges faced and how they were met.

---

[†] https://koa.ipac.caltech.edu

[‡] http://montage.ipac.caltech.edu

## 2. THE KOA VO-COMPLIANT IMAGE INTERFACE

### a. Technical Overview and Implementation Challenges

The program interface complies with version 2.0 of the Simple Image Access Protocol (SIAPv2)[§], proposed by the International Virtual Observatory Alliance (IVOA)[**], a consortium of 20 national Virtual Observatory projects that negotiate international standards for data discovery and acquisition. These standards define common sets of metadata that enable services to discover data across distributed archives in a uniform fashion. Once the services are recorded in the VO registry—a "yellow pages" for VO-services—data can be discovered by and integrated into VO-aware data integration engines that query the registry, such as Aladin at the Centre de Données astronomiques de Strasbourg (CDS); the Data Discovery Portal at the Mikulski Archive for Space Telescopes (MAST); and the Tool for Operations on Catalogues and Tables (TOPCAT). KOA responded to a request from the IVOA for evaluation and feedback of drafts of the SIAPv2 distributed in 2014 and 2015, and released in August 2015. The SIAPv2-compliant interface provides access to data acquired by all nine instruments equipped with an imaging mode. Table 1 shows these instruments in bold type: a tenth instrument, HIRES, operates purely as a spectrograph.

Briefly, the KOA program interface is implemented as a RESTful service[††] for use on the command line or within a script. It discovers public imaging data at KOA through common Applications Programming Interface (API) for all instruments. Users do need to know the details of the protocol to use the interface, which constructs a URL that submits a query to KOA and returns a list of images in an ASCII format; this list can be parsed by a program or script.

Table 1: Summary of the Keck instruments archived in KOA, their observing modes, nights archived (including full and shared nights), data volume and number of files (current as of May 15, 2016). Instruments in bold are equipped with an imaging mode.

| **Instrument** | **Modes** | **Bandpass (μm)** | **Nights** | **Raw Data Vol (TB)** | **Number of data files** |
|---|---|---|---|---|---|
| HIRES | Spectroscopy only | 0.3-1.0 | 3,131 | 5.1 | 366,203 |
| **NIRSPEC** | **Imaging**, spectroscopy | 0.95-5.5 | 1,782 | 0.8 | 530,699 |
| **NIRC2** | **Narrow-, medium, wide band imaging**; spectroscopy | 1-5 | 1,881 | 2.0 | 705,415 |
| **LRIS** | **Imaging**, spectroscopy | 0.3-10 | 3,649 | 7.2 | 472,864 |
| **MOSFIRE** | **Imaging**, spectroscopy | 1-2.4 | 538 | 2.4 | 199,028 |
| **DEIMOS** | **Imaging**, spectroscopy | 4-1.05 | 1,459 | 11.5 | 135,420 |
| **ESI** | **Imaging**, spectroscopy | 0.4-1.1 | 877 | 0.8 | 60,309 |
| **OSIRIS** | **Imaging**, spectroscopy | 1-2.5 | 941 | 3.0 | 152,342 |
| **LWS** | **Imaging**, spectroscopy | 3.5-25 | 299 | 0.06 | 27,737 |
| **NIRC** | **Imaging,** spectroscopy | 1-5 | 962 | 0.46 | 277,989 |

§ http://www.ivoa.net/documents/SIA/20151223/

** http://ivoa.net

†† https://en.wikipedia.org/wiki/Representational_state_transfer

KOA provides extensive sample queries[‡‡] and a query builder form[§§] to help users build queries that comply with the input parameters that are defined in Table 2, with those implemented by KOA shaded gray. A sample query has this structure:

```
https://koa.ipac.caltech.edu/cgi-bin/VOServ/nph-searchImage?POS=CIRCLE
150.0+45.0+5.0&INSTRUMENT=DEIMOS&TIME=2002-06-14T00:00:00/2002-06-
16T10:00:00&RESULTFORMAT=ipac
```

This query returns DEIMOS images within a 5° radius of RA =150°, Dec =45°, for a date range of 2002 June 14 to 2002 June 16, and returns results in IPAC column-delimited table format. The service operates by discovering images for a specified instrument by performing a cone, box or polygon search on the sky, and returns lists of images whose footprints on the sky intersect the area of the cone or box, with the lists filtered according to user-specified criteria of, e.g. time range. There were two major efforts needed to support the service, which are described in the next two sections:

- Managing uniform metadata tables across all instruments to comply with the protocol, and
- Mechanisms for efficient searching for, and filtering of, data without the overhead of a client-server database.

Table 2: Parameters that can be queried as defined in SIAPv2. Those implemented by KOA are shaded in grey. The "+" sign denotes the two mandatory parameters, POS (position) and instrument.

‡‡ https://koa.ipac.caltech.edu/UserGuide/program_interface.html#query_examples

§§ https://koa.ipac.caltech.edu/applications/VOServ/

| Parameter | Value |
|---|---|
| POS+ | CIRCLE (degrees) ra dec radius, POLYGON*: at least 3 vertices, RANGE*: band. |
| BAND | energy interval (m): scalar value or range of values. |
| TIME | Time interval in term of time stamps as *yyyy-mm-dd, yyyy-mm-ddThh:mm:ss.xxx* or numeric value for MJD. |
| POL | polarization state: I, Q, U, V, RR, LL, etc. |
| FOV | field of view (degree): scalar value or range of values. |
| SPATRES | Spatial resolution (arcsec/pixel): scalar value or range of values. |
| EXPTIME | Exposure time (second): scalar value or range of values. |
| ID | Identifier of data set: string-valued, case-insensitive, substring match. |
| COLLECTION | Name of data collection: string-valued. |
| FACILITY | Name of facility, usually telescope: string-valued. |
| INSTRUMENT+ | Name of instrument: string-valued. |
| DPTYPE | Type of data product: string-valued, image or cube. For KOA, image only. |
| CALIB | Calibration level (no units): non-negative integer like 0, 1, or +1, or range of non-negative integer. |
| TARGET | Name of target: string-valued. Case insensitive and substring match. |
| TIMERES | temporal resolution (second) : scalar value or range of values. |
| SPECRP | Spectral resolving power (no units): scalar value or range of values. |
| FORMAT | Data format, string-valued, FITS, JPEG, etc. For KOA, FITS only. |
| UPLOAD | Upload a table of values to be referenced by one of above parameters. |
| resultformat | Return data formats, votable, html, ipac, csv, tab, or json. |

### b. Managing metadata tables for the KOA Image Access service

Underpinning the service is a metadata catalog constructed for each instrument that contains all the information needed to support queries, including a specification of the image footprint on the sky. The KOA metadata are, however, highly heterogenous because all the Keck instruments were developed by independent teams, who never intended their data products to be ingested into a modern archive[3]. Fully one-half of our one full-time equivalent (FTE) of development effort was spent in the creation and verification of the metadata tables. In no case could all metadata simply be extracted from the database and placed into the units required by SIAPv2 as needed. Information was also extracted from the FITS file headers, and in some cases, calculated by combining information in the database, in the FITS files and in the instrument documentation. For example, fields such as time, spatial resolution and instrument field-of-view required calculation in nearly all cases.

Table 3 summarizes the wide variation in the sources used assembling the metadata of each instrument. The instruments are summarized in the columns, with each instrument designated by a two-character code. The cells are color-coded according to the origin of the metadata.

Table 3: Origin of the VO-required metadata for each of the Keck Instruments. The instruments are described in Table 1 and identified here by a mnemonic two-character code.

| Metadata Table Columns | Output Table Columns | DE | ES | LR | LW | MF | N1 | N2 | NS | OS |
|---|---|---|---|---|---|---|---|---|---|---|
| NAXIS | | ● | ● | ● | ● | ● | ● | ● | ● | ● |
| NAXIS1 | ● | ● | ● | ● | ● | ● | ● | ● | ● | ● |
| NAXIS2 | ● | ● | ● | ● | ● | ● | ● | ● | ● | ● |
| RA | ● | ● | ● | ● | ● | ● | ● | ● | ● | ● |

| **Metadata Table Columns** | **Output Table Columns** | **DE** | **ES** | **LR** | **LW** | **MF** | **N1** | **N2** | **NS** | **OS** |
|---|---|---|---|---|---|---|---|---|---|---|
| DEC | ● | ● | ● | ● | ● | ● | ● | ● | ● | ● |
| INSTRUME | ● | ● | ● | ● | ● | ● | ● | ● | ● | ● |
| ELAPTIME | ● | ● | ● | ● | ● | ● | ● | ● | ● | ● |
| TARGNAME | ● | ● | ● | ● | ● | ● | ● | ● | ● | ● |
| OBJECT | ● | | | | | | | ● | ● | ● |
| KOAID | ● | ● | ● | ● | ● | ● | ● | ● | ● | ● |
| KOAIMTYP | | ● | ● | ● | ● | ● | ● | ● | ● | ● |
| FILEHAND | | ● | ● | ● | ● | ● | ● | ● | ● | ● |
| TELESCOP | | ● | ● | ● | ● | ● | ● | ● | ● | ● |
| UTC | ● | ● | ● | ● | ● | ● | ● | ● | ● | ● |
| DATE-OBS | ● | ● | ● | ● | ● | ● | ● | ● | ● | ● |
| MJD_OBS | ● | ● | ● | ● | ● | ● | ● | ● | ● | ● |
| SPECRES | | | | | | | | ● | ● | |
| WAVEBLUE | ● | | | | | | | ● | ● | ● |
| WAVERED | ● | | | | | | | ● | ● | ● |
| FILTER | ● | ● | ● | ● | ● | ● | ● | ● | ● | ● |
| SLITNAME | ● | ● | ● | ● | ● | ● | ● | ● | ● | |
| PROGID | | ● | ● | ● | ● | ● | ● | ● | ● | ● |
| PROGINST | | ● | ● | ● | ● | ● | ● | ● | ● | ● |
| PROGPI | | ● | ● | ● | ● | ● | ● | ● | ● | ● |
| PROGTITL | | ● | ● | ● | ● | ● | ● | ● | ● | ● |
| SEMID | | ● | ● | ● | ● | ● | ● | ● | ● | ● |
| cntr | | | | | | ● | | ● | | ● |
| cra | | | | | | ● | | ● | | ● |
| cdec | | | | | | ● | | ● | | ● |
| CTYPE1 | | | | | | ● | | ● | | ● |
| CTYPE2 | | | | | | ● | | ● | | ● |
| CRPIX1 | | | | | | ● | | ● | | ● |
| CRPIX2 | | | | | | ● | | ● | | ● |
| CRVAL1 | | | | | | ● | | ● | | ● |
| CRVAL2 | | | | | | ● | | ● | | ● |
| CDELT1+ | | | | | | ● | | ● | | ● |
| CDELT2+ | | | | | | ● | | ● | | ● |
| CROTA2+ | ● | | | | | ● | | ● | | ● |
| EQUINOX | | | | | | ● | | ● | | ● |
| ra1 | | | | | | ● | | ● | | ● |
| dec1 | | | | | | ● | | ● | | ● |
| ra2 | | | | | | ● | | ● | | ● |
| dec2 | | | | | | ● | | ● | | ● |
| ra3 | | | | | | ● | | ● | | ● |
| dec3 | | | | | | ● | | ● | | ● |
| ra4 | | | | | | ● | | ● | | ● |
| dec4 | | | | | | ● | | ● | | ● |
| hdu | | | | | | ● | | ● | | ● |
| size | ● | ● | ● | ● | ● | ● | ● | ● | ● | ● |
| TIME | | ● | ● | ● | ● | ● | ● | ● | ● | ● |
| SPATRES | ● | ● | ● | ● | ● | ● | ● | ● | ● | ● |

| **Metadata Table Columns** | **Output Table Columns** | **DE** | **ES** | **LR** | **LW** | **MF** | **N1** | **N2** | **NS** | **OS** |
|---|---|---|---|---|---|---|---|---|---|---|
| FOV1 | | ● | ● | ● | ● | ● | ● | ● | ● | ● |
| DPTYPE1 | ● | ● | ● | ● | ● | ● | ● | ● | ● | ● |
| CALIB | ● | ● | ● | ● | ● | ● | ● | ● | ● | ● |
| FACILITY | ● | ● | ● | ● | ● | ● | ● | ● | ● | ● |
| | | | | | | | | | | |

| | |
|---|---|
| (grey) | Columns included in the output |
| (white) | Columns included in the metadata table |
| (cyan) | Data from database |
| (yellow) | Data from FITS header |
| (magenta) | Derived data |

Because the SIAPv2 performs spatial searches, the World Coordinate System (WCS) keywords describing the image footprint on the sky—the parameters CTYPE1 through CROTA2 in Table 3—are of particular importance. Only the observations of OSIRIS, NIRC2 and MOSFIRE record this WCS information in the FITS headers. For the remaining instruments, there is insufficient information to compute the WCS headers consistently and reliably and the fields-of-view are too small for services such as `astrometry.net`[***] to return astrometric calibrations. In those cases, we assume the image is centered on the position on the sky where the telescope was pointing. We convert polygon searches to cone searches by enclosing the requested search area in a circular region. For both cone and polygon searches, we then use the instrument field-of-view to extend the circular region on the sky, as shown in the diagrams below for a cone search (Figure 4a) and for a polygon search (Figure 4b). This technique, in effect, pads the search areas a few arc seconds, and there is a small probability the service returns additional records over and above what would be expected from an exact search.

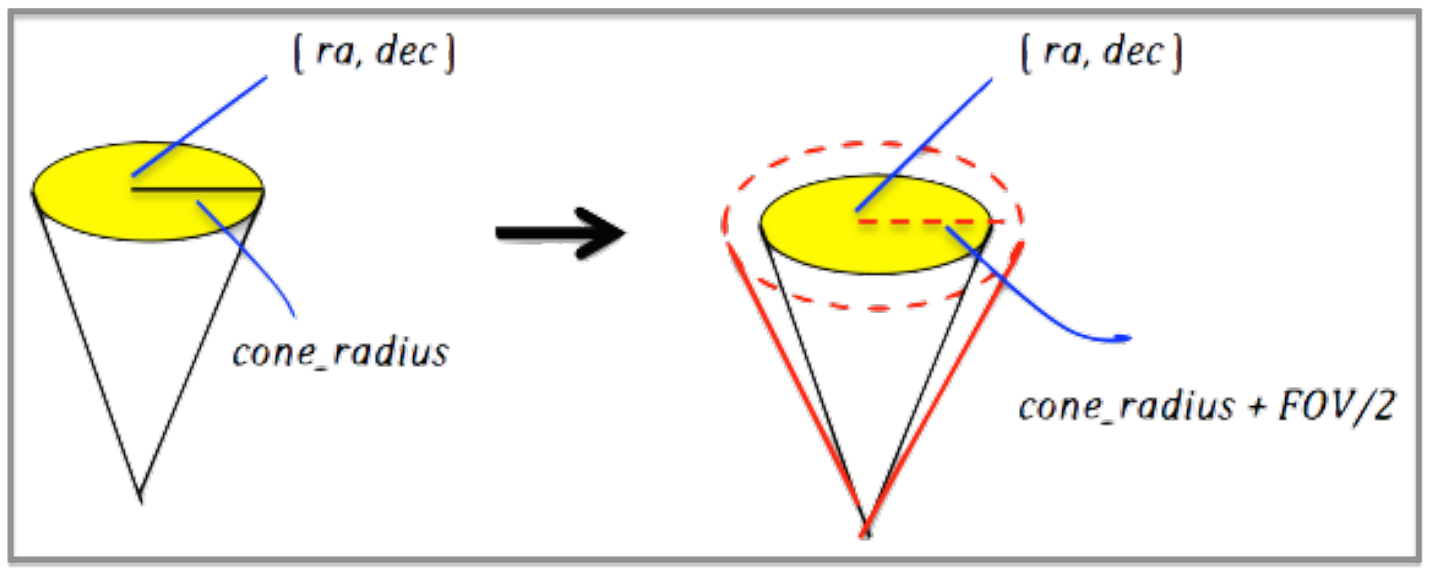

Figure 4a: Schematic showing how cone searches are managed where the WCS information is unavailable.

[***] http://astrometry.net/

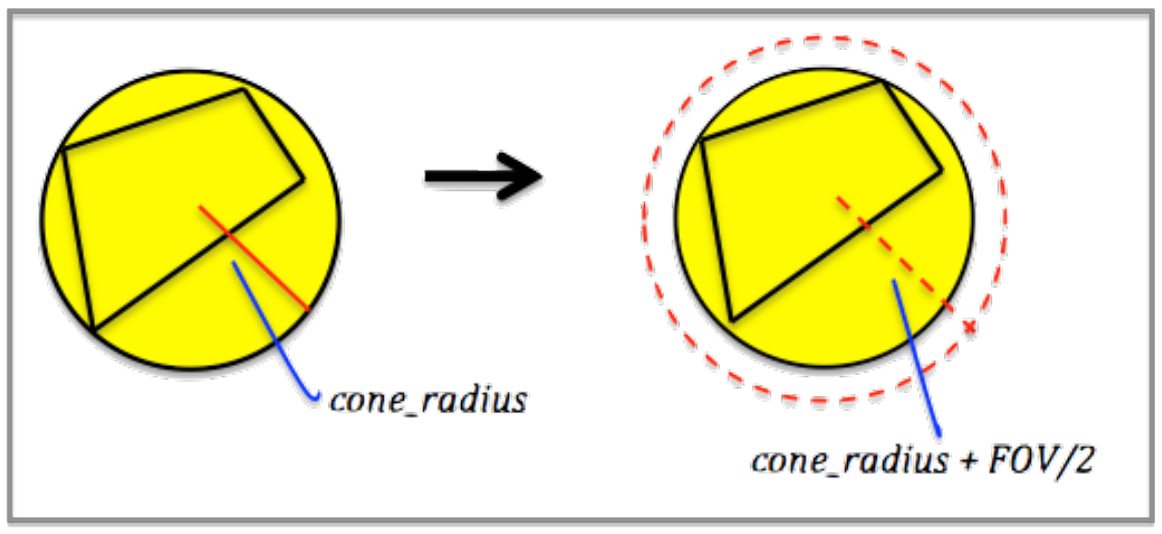

Figure 4b: Schematic showing how polygon searches are managed where the WCS information is unavailable

**c. Spatial Searches and Filtering of Results**

One of our goals was to use freely available software to build a low-maintenance service. In particular, we wished to avoid the overhead in installing and administering a client-server database. Our spatial searches have therefore been optimized through the use of R-tree indices[4], which can be implemented without the use of such a client-server database, with an embedded Open Source database, SQLite[†††], to perform post-spatial-query filtering as needed.

R-trees are tree data structures used to access multi-dimensional information such as geographical coordinates, or, in our case, images. One of the virtues of R-trees is they are applicable to images that are densely and sparsely distributed images, and to point sources as well. They are in wide use, for example, in popular applications that search for restaurants or landmarks. R-trees operate by grouping nearby objects, overlapping as needed, and represent them with their minimum-bounding rectangle in the next higher level of the tree. Figure 5 shows an example of an R-tree.

[†††] http:///www.sqlite.org

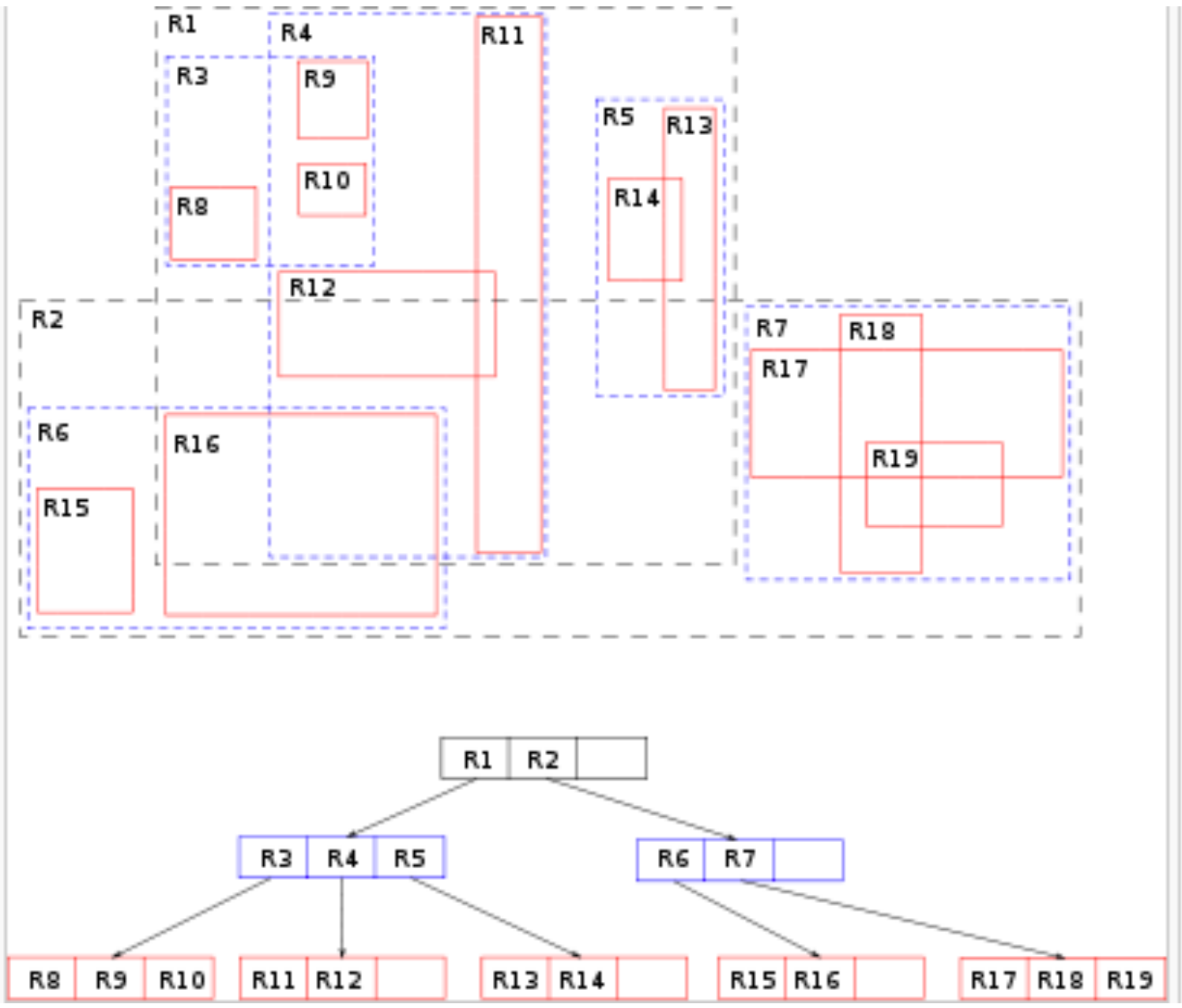

Figure 5: Example of an R-tree for 2D rectangles (from https://en.wikipedia.org/wiki/R-tree; public domain image)

In our case, the four corners of the images for the OSIRIS, NIRC2 and MOSFIRE instruments are the minimum-bounding rectangles; the four corners are represented by fields *ra1* through *dec4 i*n Table 3 and computed from the WCS fields. For the remaining instruments, the images are indexed as point locations, and then padded by the radius of the field-of-view, and are therefore effectively treated as circular objects. In both sets of cases, a search works its way down the tree, pruning branches that do not overlap with the search region. Our implementation makes use of memory-mapped files, which map a file on disk to a range of addresses within an application's address space. The application can then access files on disk in the same way it accesses dynamic memory. This in turn makes file I/O much faster than opening a file and transferring its contents to memory. Once the tree is built, which can take hours or days for very large tables, billion-record tables can be searched in milliseconds. In the case of KOA, the image metadata tables contain at most only several hundred thousand records, as shown in column 6 of Table 1, and building the R-tree indices only takes seconds. Thus, as the metadata tables are updated each day as new data are released, the indices are easy to rebuild with no impact on performance.

This indexing method is itself based on a prototype developed for the Virtual Astronomical Observatory (VAO)[5], and was developed further in Montage as way of discovering images over wide spatial areas that will be input for the creation of mosaics. When fully documented, it will be released as the module `mSearch` as part of the Montage distribution. It has already been implemented within the Infrared Processing and Analysis Center (IPAC) to underpin searches for images released by the Spitzer Space Telescope, the Wide-field Infrared Survey Explorer (WISE) and the NASA Exoplanet Archive.

Post-query filtering is performed with the SQLite relational database management system. SQLite is not a client-server database, but rather an embedded database that is linked to the application and becomes part of the application program. It is available as a file for download, and the "vanilla" version suitable for our application is free of charge. It is in wide commercial use: for example, it is built in to cell phone operating systems, where it is used instead of a client-server database to extend battery life and reduce data charges. Thus, we have been able to implement the VO service with freely available software that avoids the overhead involved in using a client-server database.

# 3. THE KOA MOVING OBJECT SEARCH SERVICE

KOA archives over 20,000 observations of the major Solar System planets, as well as over 18,000 observations of Pluto and its satellites, many of which were acquired to support the New Horizons encounter. Consequently, we are developing the Moving Object Search Service (MOSS), which, when released in late 2016, will return lists of images and spectra containing solar system targets, according to the input specification from a web form of target, time range and tolerance in search radius. The development of this service exploits and extends the expertise and tools used in the development of the VO interface, especially the R-tree indexing and post-query filtering, and takes advantage of the Spacecraft Planet Instrument Camera-matrix Events (SPICE) information system[‡‡‡] released by the NASA Navigation and Ancillary Information Facility (NAIF)[§§§] to compute the track of a target across the sky.

Figure 6 shows a map of the ephemeris of Pallas over the operational lifetime of the Keck Observatory, and the public NIRC2 observations of Pallas available in KOA. This plot was created with the Montage tool `mViewer`[****], which creates images with labels and overlays from command line input and released as part of the version 4.x distribution[††††]. The rest of this section describes how these matches are derived.

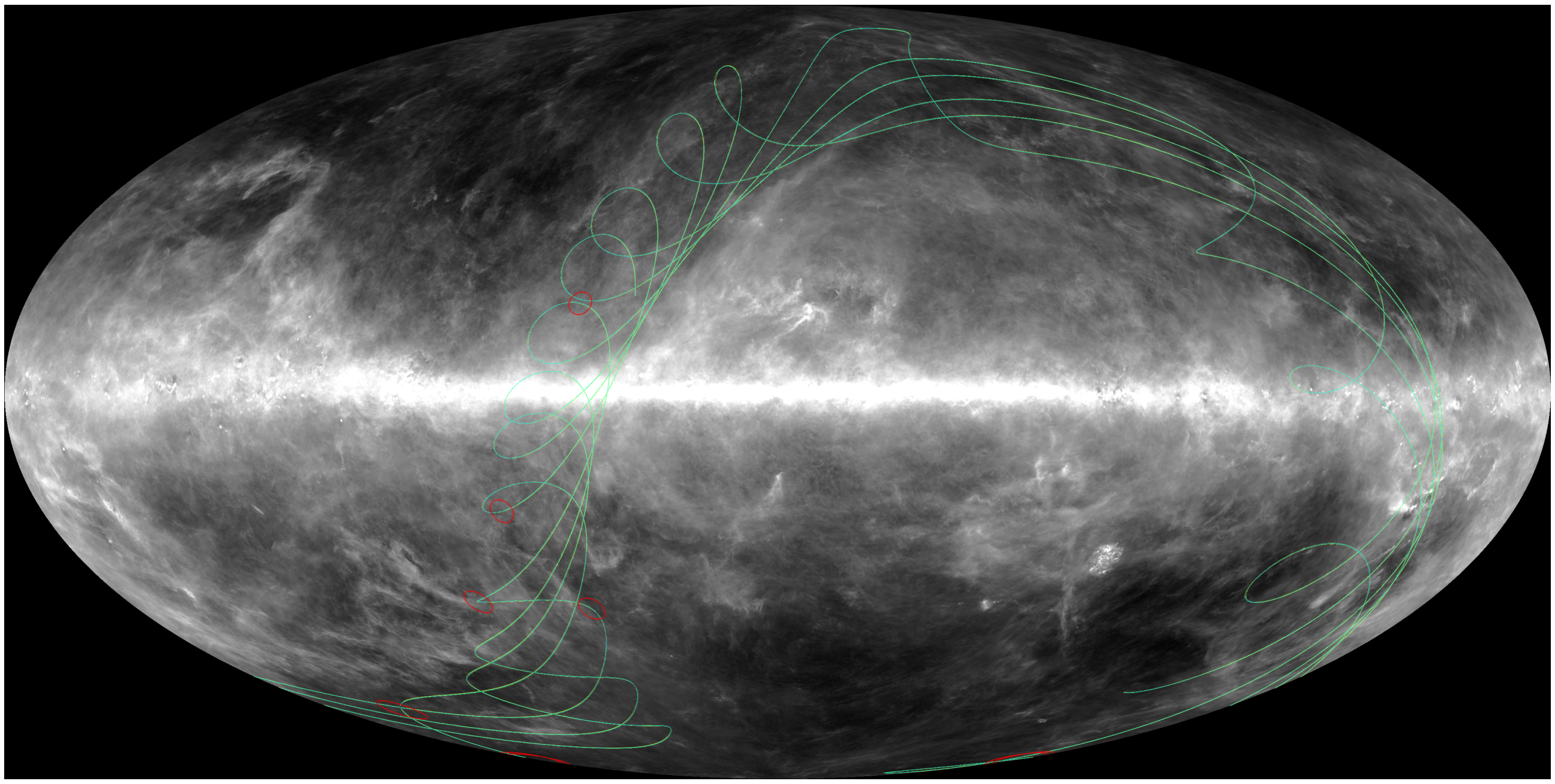

Figure 6: Map in Galactic Coordinates of the ephemeris of Pallas (green track) computed from 1995-01-01 to 2015-10-10, with the matching observations for NIRC2 superposed (red circles). The underlying all-sky image is a 100 μm map as measured by the Infrared Astronomical Satellite (IRAS).

‡‡‡ https://naif.jpl.nasa.gov/naif/spiceconcept.html
§§§ https://naif.jpl.nasa.gov/naif/index.html
**** http://montage.ipac.caltech.edu/docs/mViewer.html
†††† http://montage.ipac.caltech.edu/docs/download2.html

## a. Technical Overview of the Moving Target Search.

The service operates in a fashion that is analogous to the VO service, as follows:

- Build metadata tables for raw public images and spectra for each instrument. Spectra are regarded as "skinny images" for this purpose.
- Build R-tree indices for each instrument. As the service performs searches in space and time, so the indices are built in space and time.
- Update the tables and the indices each day, as new observations are made public.
- With these underpinnings in place, searches for moving targets are performed as follows:
    - Use the SPK library, a component of SPICE, to create an ephemeris table for the input target and time range.
    - Create bounding boxes around groups of ephemeris points.
    - Search for data in the bounding boxes using the R-trees to find candidate records.
    - Post-filter the candidates to find the "hits" and "near-misses."
- Create graphics of the ephemeris track, bounding boxes and target matches.

## b. The Moving Object Search Metadata Tables.

Table 4 shows the schema for metadata tables to support searches for imaging and spectral data for, with example values for one instrument, NIRC2. As with the VO table, some of these values are accessed from the database or from the FITS files, while others are calculated. The first release will use separate tables for the VO and MOSS services, but the long-term plan is to combine them. Because the size and orientation of the spectroscopic slit on the sky is generally not known, we treat the spectra as point sources and perform cone searches, in the same fashion as for searches for images in the VO service.

Table 4: Example Moving Object Service metadata table schema for the NIRC2 instrument.

| Metadata Field | Description | Example (NIRC2) | Origin (NIRC2) |
|---|---|---|---|
| KOAID | Unique file identifier assigned by KOA | N2.20121202.14856.fits | Database |
| date_obs | Date of observation | 2012-12-02 | Database |
| UT | Universal Time | 04:09:31.72 | Database |
| mjd_obs | Start time of observation | 04:90:29.35 | Database |
| Elaptime | Duration of observation | 20.00 | Database |
| RA | Right Ascension (J2000) | 355.1024000000 | Database |
| Declination | Declination (J2000) | 44.3338600000 | Database |
| Filehand | File handle of observation | /koadata10/NIRC2/20121202/lev0/N2.20121202.14856.fits | Database |
| imgtyp | Classification of file | Science | Database |
| Mode | Image or spectral mode | Image | Database |
| Field-of-view | Field of view for each instrument mode | 10 arcsec – narrow mode<br>20 arcsec – medium mode<br>40 arcsec – wide mode | Calculated |

### c. Creation of R-tree indices and Operation of the Search.

R-trees are well suited to indexing in multiple dimensions and are created on space and time for the MOSS with `mMovingTarget`, an extension of `mSearch`, to three or more dimensions that will be deployed in a future release of Montage. It is quite possible to use a two-dimensional spatial index followed by filtering of images in time, an approach taken by the Infrared Science Archive (IRSA) Moving Object Search Tool (MOST)[‡‡‡‡] for searches of the WISE image surveys. There are, however, two advantages to using three-dimensional R-trees instead. One is that the code is easier to maintain. The other is that the search is much faster. A search for the Jovian satellite Io over the full five years of WISE surveys is 400 times faster with three-dimensional indices, simply because the many image footprints that intersect the path of Io do not have to be searched individually for matches in time.

The schematic in Figure 7 shows how the MOSS operates by using an example of searching for observations of Io over a period of three years:

- Use the SPICE SPK library to retrieve the latest orbital elements for the target and tabulate an ephemeris with time steps of 75 minutes. We compute the target's sky position at each MJD mid-point. The table contains (RA, Dec, MJD, duration of observation).
- Use the ephemeris table and create from it a set of bounding boxes that will be used for the search. The time interval for computing the box is 15 hours, and its width is determined by using the `bnd` library in Montage to create a minimum bounding box that encloses the points, and padded according to the value of the near-miss tolerance.
- The ephemeris boxes are used to search the R-tree indices and identify candidates that are recorded in a match table.
- For the match table, compute the distance between each candidate points and the ephemeris positions and filter out those that are outside the maximum tolerances. The final set of matches and the near misses are returned to the user.

[‡‡‡‡] http://irsa.ipac.caltech.edu/applications/MOST/

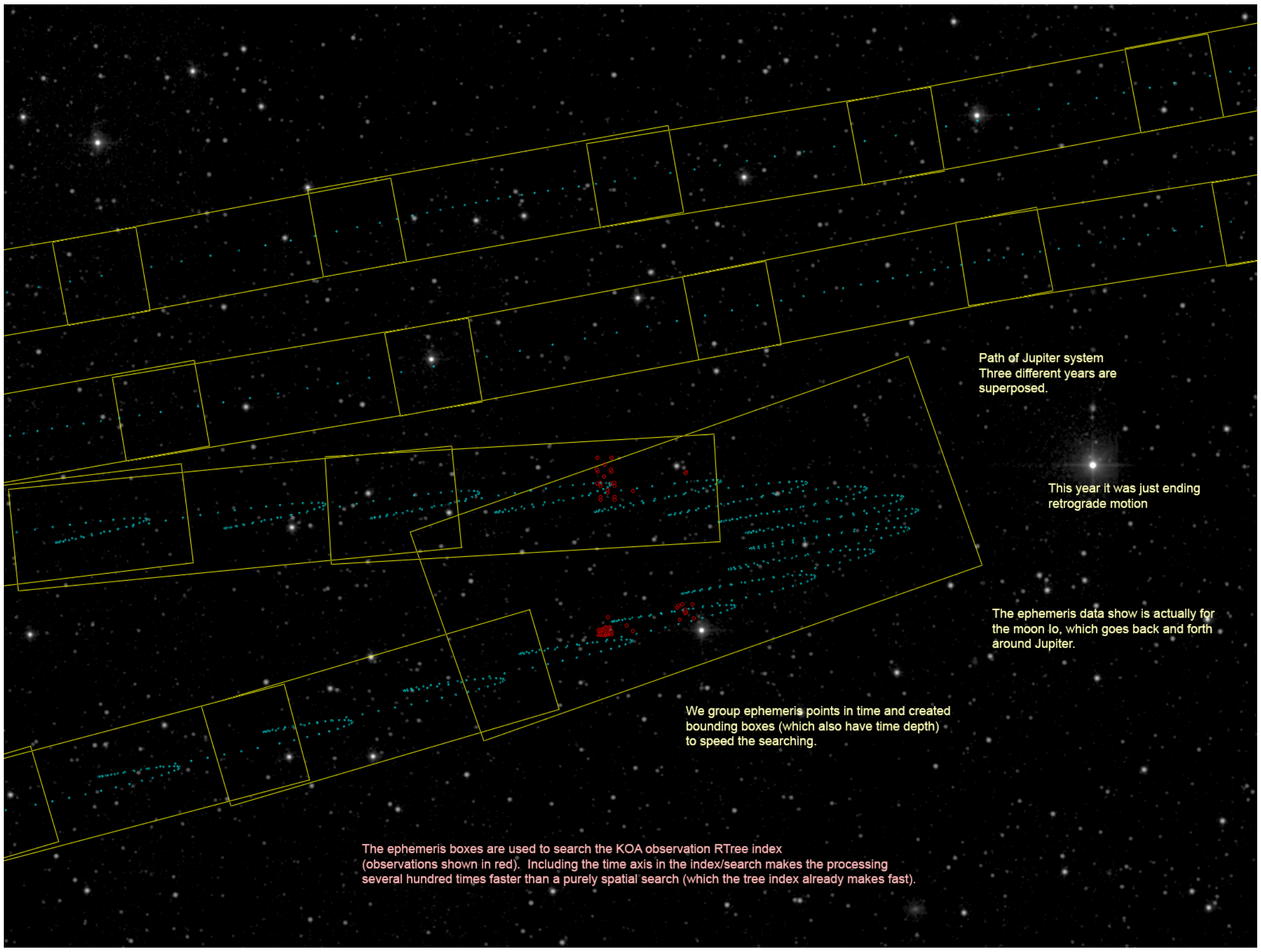

Figure 7: The Moving Object Search in action, as shown by a three year ephemeris for Io (blue). The bounding boxes are shown in yellow, and matches with NIRC2 images are shown in red.

A graphical representation of the results is clearly valuable. Yet as Figures 6 and 7 show, the matches are scattered in space (as well as time), and so the graphics are created with a heuristic approach that plots clusters of matched points. Figure 8 shows 51 matches for Pallas centered at RA= $18^h\ 03^m$ and Dec= 17° 10′. The ephemeris table was subset by creating R-tree indices to filter those points that lay in the plot region.

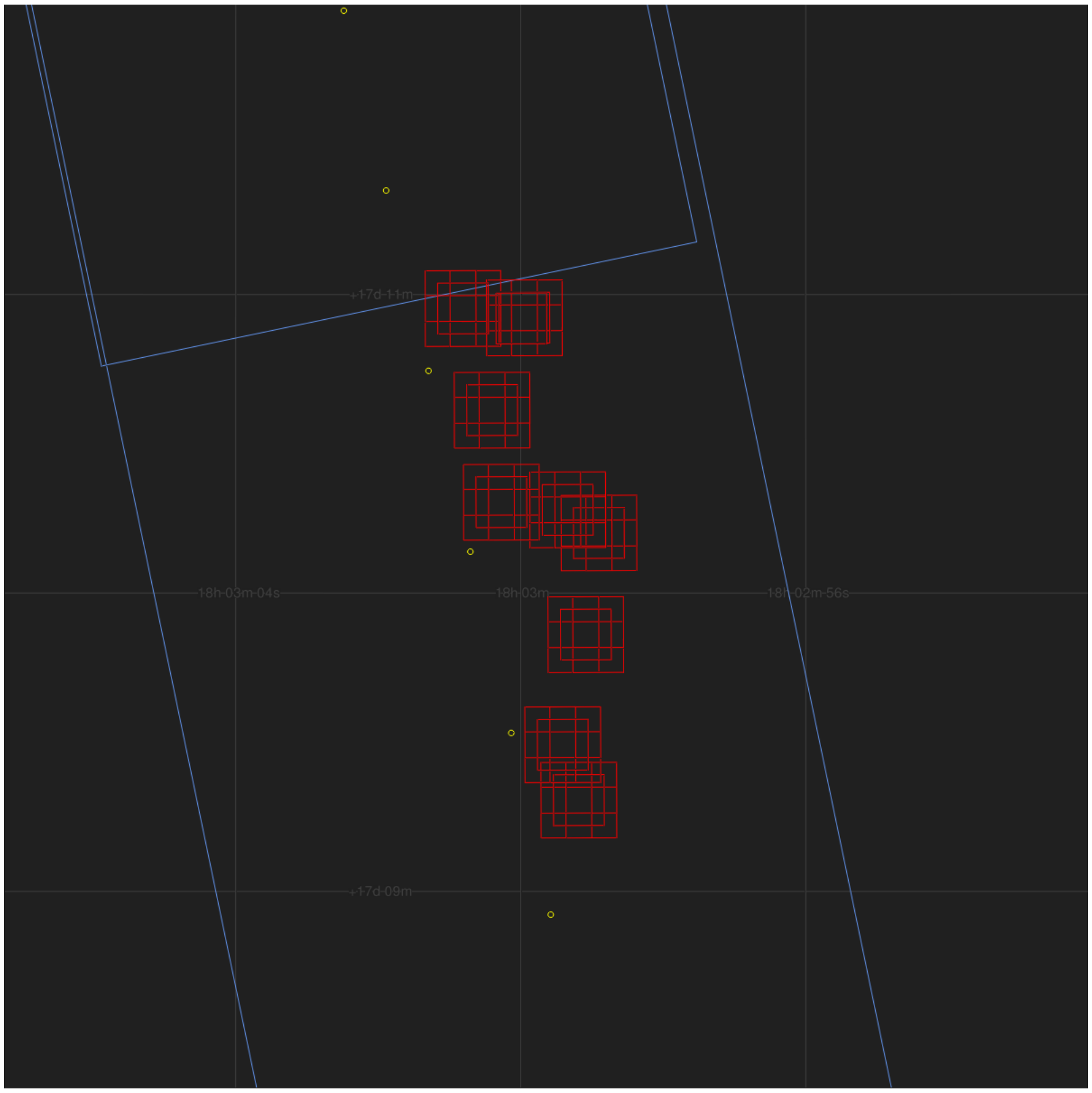

Figure 8: Close-up of 51 matches of Pallas with NIRC2 images, centered at RA= $18^h\,03^m$ and Dec= 17° 10′. Blue boxes: outlines of bounding boxes; yellow dots: ephemeris points; red boxes: NIRC2 image outlines

## d. Performance

Table 5 summarizes the performance of the Moving Object Search service for the 20-year search for matches of Pallas with NIRC2 images. The generation of the ephemeris is limits the performance, but the time required scales with the interval of the search: a one-year time interval would require less than 1 second for the calculation of the ephemeris. The generation of the graphics will, in most cases, limit the performance.

Table 5: Performance of the Moving Object Search for Pallas over a 20-year period.

| **Task** | **Time(s)** |
| --- | --- |
| ***Moving Target Searches*** | |
| Get ephemeris of target | 0.1 |
| Generate ephemeris points | 17 |
| Generate candidate matches | 0.2 |
| Post filtering | 0.04 |
| ***Graphics Generation*** | |
| Get matches from table | < 0.1 |
| Create R-tree indices for subsetting ephemeris table | 22 |
| Generate JPG | 11 |

# 4. CONCLUSIONS

There have been two major challenges in successfully developing the VO service and the MOSS. One is creating uniform and consistent metadata to support queries. In those cases where there was insufficient information to compute the required metadata, we were able to make approximations that did not place any compromises on usability and quality of results. Managing metadata is time-consuming when the data set is inhomogeneous: in the case of the VO-service, this task consumed half the development time.

The second challenge was creating a low-maintenance, low-cost method of supporting searches. The use of R-tree indices, coupled with the SQLite embedded database, achieved this goal. This methodology will have general applicability, such as underpinning a research team's archive. When complete, all the code will be made public as part of the Montage distribution and accessible through GitHub.

**Acknowledgements:** The Keck Observatory Archive (KOA) is collaboration between the W. M. Keck Observatory (WMKO) and the NASA Exoplanet Science Institute (NExScI). Funding for KOA is provided by the National Aeronautics and Space Administration (NASA). WMKO is operated as a scientific partnership among the California Institute of Technology, the University of California, and NASA. The Observatory was made possible by the generous financial support of the W. M. Keck Foundation. NExScI is sponsored by NASA's Origins Theme and Exoplanet Exploration Program, and operated by the California Institute of Technology in coordination with the Jet Propulsion Laboratory. This material is based in part upon work supported by NASA under Grant and Cooperative agreement No. NNX13AH26A. This work made use of Montage. It is funded by the National Science Foundation under Grant Number ACI-1440620, and was previously funded by the National Aeronautics and Space Administration's Earth Science Technology Office, Computation Technologies Project, under Cooperative Agreement Number NCC5-626 between NASA and the California Institute of Technology.